\documentclass[conference]{IEEEtran}
\IEEEoverridecommandlockouts
\usepackage{cite}
\usepackage{amsmath,amssymb,amsfonts}
\usepackage{algorithmic}
\usepackage{graphicx}
\usepackage{textcomp}
\usepackage{xcolor}

\usepackage{booktabs} % For formal tables
\usepackage{subfig}
\usepackage{hyperref}
\usepackage{todonotes}
\usepackage{color,soul}
\usepackage{listings}
\usepackage[framemethod=tikz]{mdframed}
\usetikzlibrary{shadows}

\lstdefinelanguage{mylang}{
  alsoletter={*,", 10, >=, :, 74,40,-},
  keywords=[1]{producer,NOTIFY_INTEREST,NOTIFY_DATA,STORE_FUNCTION,topo1,START_FUNCTION},
  keywords=[2]{"Drone","LiDAR","Li*","post_processing_func", "IF, RESULT, 10 , >=, ","lat:40*", "long:-74*"},
  keywords=[3]{new},
}

\def\BibTeX{{\rm B\kern-.05em{\sc i\kern-.025em b}\kern-.08em
    T\kern-.1667em\lower.7ex\hbox{E}\kern-.125emX}}
\begin{document}

%\title{A Hybrid Edge Cloud Framework For IoT Based Data-Driven Pipelines}
\title{Edge Based Data-Driven Pipelines (Technical Report)}

\author{\IEEEauthorblockN{Eduard Gibert Renart}
\IEEEauthorblockA{\textit{Rutgers Discovery Informatics Institute}\\
\textit{Rutgers University}\\
Piscataway, NJ, 08854, USA\\
egr33@rutgers.edu
}
\and
\IEEEauthorblockN{Daniel Balouek-Thomert}
\IEEEauthorblockA{\textit{Rutgers Discovery Informatics Institute}\\
\textit{Rutgers University}\\
Piscataway, NJ, 08854, USA \\
}
\and
\IEEEauthorblockN{Manish Parashar}
\IEEEauthorblockA{\textit{Rutgers Discovery Informatics Institute} \\
\textit{Rutgers University}\\
Piscataway, NJ, 08854, USA \\
}
}

\maketitle

\begin{abstract}
This research reports investigates an edge on-device stream processing platform, which extends the serverless computing model to the edge to help facilitate real-time data analytics across the cloud and edge in a uniform manner. We investigate associated use cases and architectural design. We deployed and tested our system on edge devices (Raspberry Pi and Android Phone), which proves that stream processing analytics can be performed at the edge of the network with single board computers in a real-time fashion. 
\end{abstract}


\section{Introduction}
The proliferation of the Internet of Things (IoT) paradigm and the exponential growth of connected devices has the potential of transforming a wide range of applications, impacting science, engineering, medicine, business, and society in general. At the same time, the data resulting from these connected devices presents new challenges that have to be addressed before this potential can be effectively realized. In addition to Doug Laney's initial 3 V's big data challenges (Volume, Velocity, and Variety)~\cite{3v}, IoT data analytics presents new challenges such as:    
\begin{itemize}
\item Processing machine-generated data streams produced at high-speed rates (thousands of messages/sec) compared to big data which is mostly human-generated data.  

\item Managing sensors and actuators runtime parameters such as location and accuracy over the lifecycle of an application.    

\item Identifying context-sensitive and critical insights using data produced in a geographically distributed manner while dealing with minimal latencies.  

\item Enabling flexible abstractions and logic to allow applications to efficiently query and store data products in a timely manner.
\end{itemize}

Major stream processing engines use cloud-centered architecture, in which data streams from IoT devices are sent to the cloud to be processed. Processing all data in such a fashion introduces high bandwidth cost between the edge and the cloud, and increases latency and response time of applications, preventing fast insights closer to where the data originates. As an example, commercial jets generate 10 TB of data for every 30 minutes of flight~\cite{cisco}. This makes impractical to transport all of that data to the cloud, simply because it doesn't make sense to waste bandwidth on transporting the data to the cloud when the majority of the data will not be useful. As the number of IoT connected devices grows, it is predicted that by 2020 there will be 50 to 100 billion IoT devices connected to the Internet~\cite{McAuley}, making the cloud-centric model unsustainable and unable to match expectations of reactivity and flexibility. 

In this context, edge computing introduces the idea of pushing intelligence and processing capabilities closer to the physical assets and sensors ~\cite{edgecomputing}~\cite{edge2}~\cite{edge3}. This approach requires leveraging resources that may not be continuously connected to a network along with the ability for applications and third-party systems to store and query data products, rather then devices, in an efficient manner.

Performing edge analytics presents several technical challenges at different levels. Sensors and actuators need a programming abstraction to enable data streaming and discovery of services without knowledge of data consumers in a distributed computing environment. Applications and developers require a programming abstraction to perform processing based on the content of the data with the ability to trigger streaming topology and workflows seamlessly between the edge and the core of the network. Also, such features need to be implemented in a unified lightweight software stack to allow a deployment both on limited (edge) and robust (core) computational resources while ensuring high-throughput and scalability system-wide.  

The contribution of this paper is a novel software stack, R-Pulsar, designed as a lightweight framework for IoT data analytics as a service. The main features of R-Pulsar are:  
\begin{itemize}
\item Content-based decoupled interactions between data producers and consumers to enable programmable reactive behaviors.
\item A memory-mapped data processing layer to ensure high-performance stream analytics on single-board computers.
\item An event-driven programming abstraction that resource- and location-aware operators can be deployed on in a standalone fashion or on top of existing solutions.
\item An extension of the traditional stream processing model that switches the current view of cloud-centric analytics into more distributed edge and core real-time data analytics.  
\end{itemize}

Using R-Pulsar, data streams are defined with function and resource profiles. The function profile is responsible for encapsulating the process to be executed and provide the end user with fine-grained control over deployment and execution based on location and QoS requirements. The resource profile is responsible for overcoming the limitations of the traditional Pub/Sub subscription model~\cite{mqtt} and enabling users with the ability to discover IoT resources and services at runtime. Finally R-Pulsar introduces a rule-based API that can be incorporated into the user-defined data analytics logic to perform decisions and trigger new functions on demand. 

We demonstrate the effectiveness of R-Pulsar by means of two experiments: first, we evaluate the performance and scalability of messaging, query, and storage layers. Second, we compare the response time of different engines on a complete data-processing pipeline using a real life use case. 

The rest of the paper is structured as follows: Section~\ref{sec:background} describes the common requirements for data-driven application and describes our motivating use case. Section~\ref{sec:related} presents the related and limitations of current approaches. Section~\ref{sec:framework} presents the design and implementation. Section~\ref{sec:evaluation} presents an experimental evaluation of R-Pulsar on different platforms. Section~\ref{sec:conclusion} concludes the work and suggests some future work for this paper. 

\section{Disaster Recovery Motivating Usecase}\label{sec:background} 
Emerging applications require fast analytics and low latencies. Deploying all computing tasks on the cloud has proven to be effective for data processing since the computing power on the cloud outperforms the computing capabilities of the edge. However, the bandwidth of our networks has not increased as fast as the computational power, and with the increase of data being generated at the edge, the bandwidth of the network is becoming the bottleneck. For example, Intel has estimated that a single smart car will generate 4 terabytes of data every day, and it requires data to be processed in real time in order for the vehicle to make decisions~\cite{intel}. If all of these data needed to be sent to the cloud for processing, current network bandwidths would not be capable of supporting a large number of vehicles in one area. In this case, data need to be processed at the edge for minimizing the response time and to reduce network pressure.
\\
In this context, this work focuses on a disaster recovery workflow. 
Disaster management is a process that involves four phases: mitigation, preparedness, response, and recovery. Mitigation efforts attempt to prevent hazards from developing into disasters altogether or to reduce the effects of disasters when they occur. In the preparedness phase, emergency managers develop plans of action when the disaster strikes and analyze and manage required resources. The response phase executes the action plans, which includes the mobilization of the necessary emergency services and the dispatch of first responders and other material resources in the disaster area. Finally, the aim of the recovery phase is to restore the affected area to its previous state. 

This paper focuses on the response phase and uses a multi-stage response workflow that will be deployed between the edge and at the core of the network. We start by capturing real-time data from multiple affected zones (e.g LiDAR, photogrammetry, etc.) by using multiple drones equipped with a LiDAR camera and a Raspberry Pi to perform the preprocessing stage at the edge of the network. Once the image has been pre-processed, a decision will be performed based on the content of the pre-processed data to determine if we need any further post-processing. If further processing is needed, data will either be sent to the cloud to perform a change detection with previously recorded historical data, store data into the cloud for historical data, or ask an agency to determine if building conditions are safe. To simulate this workflow we used real LiDAR images that were taken right after Hurricane Sandy struck back in 2012 in the NY and Long Island area, with a total of 741 images and 3.7 GB in size, with the biggest image size of 33.8 MB, and the smallest of 1.8 KB. For historical data we used a bigger data set of pre-Hurricane Sandy. 

The analysis of this motivating use case presents some requirements.   
\begin{itemize}
  \item Data is produced in a geographically distributed fashion.  
   
  \item There is a need for low latencies to enable fast decisions.  
  
  \item Prohibitive cost of moving all data to the core for analysis. 
\end{itemize}

The presented use case demonstrates the need for consolidating cloud- and edge-based data analytics techniques. To enable fast decisions we need to deploy low-latency algorithms at the edge of the network to pre-process the data in real time. Occasionally, due to low compute resources and the lack of historical data at the edge, it requires us to send images back to the cloud for post-processing. This constitutes the primary motivations of R-Pulsar software stack. R-Pulsar was designed to programmatically unify and abstract the core and the edge resources and to perform stream processing at the most appropriate location of the network by considering descriptive properties and latency needs. 

\begin{figure*}[htb!]
  \centering
    \includegraphics[width=1.9\columnwidth]{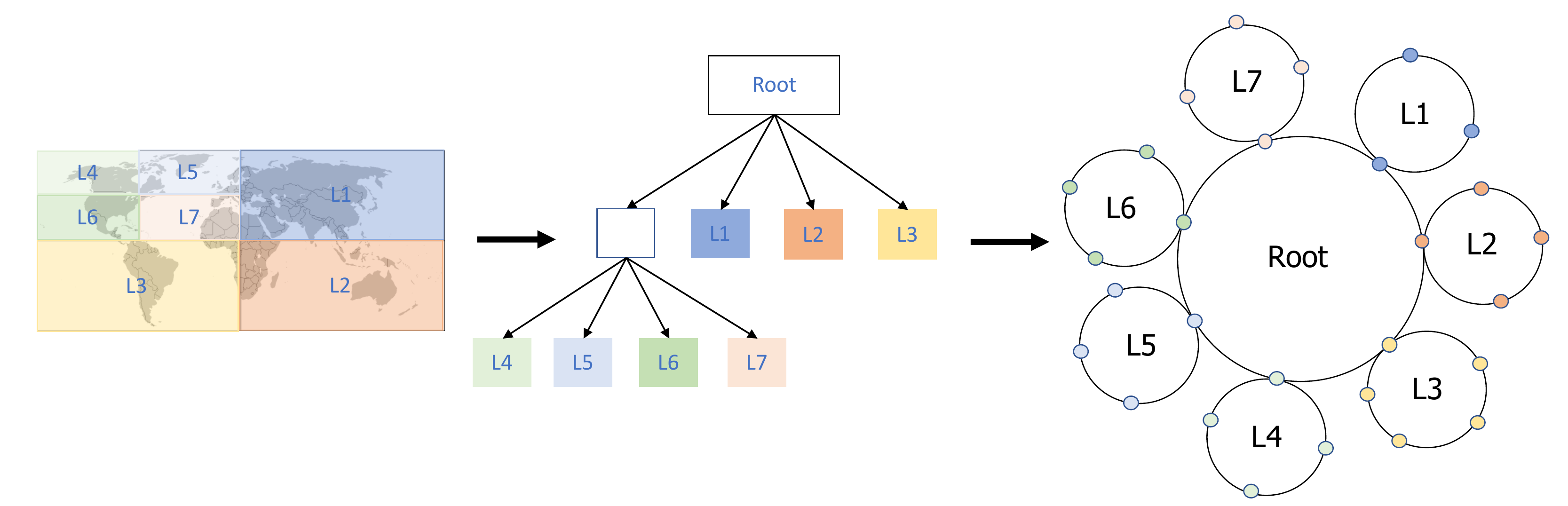}
  \caption{R-Pulsar quadtree geographical organisazation to multi layer P2P overlay network.} \label{fig:quadtree}
\end{figure*}

\section{Related Work}\label{sec:related}
\textbf{Core analytics}
has been well studied to enable real-time stream processing in the cloud. Current IoT and non-IoT data analytics systems employ a cloud-centered architecture, in which data streams from IoT devices are sent to the cloud via a gateway and processed by using distributed streaming computing technologies. Systems like Heron~\cite{heron}, Storm~\cite{storm}, Flink~\cite{flink}, Mill-Wheel~\cite{millwheel}, and Spark~\cite{spark} were developed to handle massive amounts of data at high speeds. While these approaches address scalability issues, edge-specific features such as locality awareness are not considered, which is crucial for achieving low latencies in real-time analytics. A new paradigm for deploying cloud analytics is emerging to address these features: serverless computing. The serverless computing model is a cloud-computing paradigm in which the cloud provider dynamically manages the allocation of machine resources, allowing developers to build and run applications and services without thinking about servers. Serverless applications don't require the developper to provision, scale, or manage the servers. A major drawback of serverless computing remains that it is designed to be executed at the core of the network~\cite{serverless}.

\textbf{Edge analytics:}
Fewer research efforts have been made to develop novel architectures for data analytics platforms in the edge. Mosquitto~\cite{mosquitto} is a required piece for realizing a data analytics application. Mosquitto is an open source lightweight pub/sub messaging system that is designed to be deployed on constrained devices. The main limitations of Mosquitto are: the lack of any failover or high availability mechanisms, and it was not designed to scale due to the use of bridging (a bridge lets you connect two MQTT brokers together). Some industrial systems have been implemented to explore edge analytics, such as AWS IoT~\cite{AWS-iot} or Azure IoT~\cite{azure-iot}, however, they don't offer the ability to deploy functions at the edge on demand–everything has to be deployed through the cloud. In addition, they are not built to perform high-performance analytics at the edge. Another key piece for realizing IoT edge analytics is adding location awareness into the pub/sub messaging system. Some work has been done in this area, in particular, Yuuichi et al~\cite{pub-loc} presented a location-aware topic pub/sub messaging system. The main drawback is that they rely on the topic-based pub/sub model which has already been identified as not suitable for IoT since there is no publisher or topic discovery mechanisms~\cite{mqtt}. R-Pulsar overcomes the location awareness challenge by introducing a new location-aware overlay. It eliminates the topic-based pub/sub model and replaces it with the Associative Rendezvous model.

R-Pulsar provides a full-stack platform for real-time data analytics across cloud and edge resources in a homogeneous manner. The main goal of R-Pulsar is to facilitate and automate the management of the underlying resources (edge and core)  in order to can achieve an optimal placement for the analytics functions. This approach enables combining the benefits of the edge resources (lower response time) with core resources (high compute power and storage). R-Pulsar extends the Serverless model to the edge of the network to facilitate the management of the underlying edge and core resources. R-Pulsar differentiates itself on three major aspects (i) the pub/sub queueing system is memory mapped instead of heavily relying on the filesystem for storing and caching messages (as done in Apache Kafka~\cite{kafka}) (ii) the serving layer capabilities are present within the pub/sub messaging system by integrating a lightweight SQL engine. (iii) The extension to the traditional Serverless computing model to support edge devices. 

\begin{figure*}[htb!]
\begin{center}
\subfloat[]{\includegraphics[width=0.98\columnwidth]{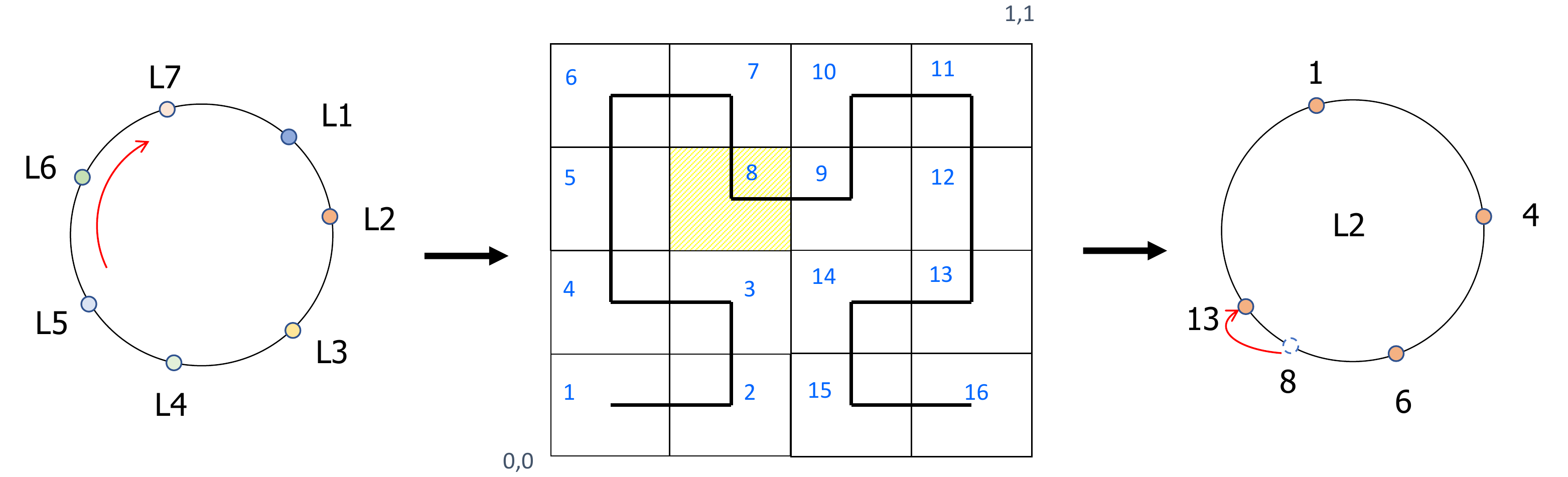}
\label{fig:single}}
\subfloat[]{\includegraphics[width=0.98\columnwidth]{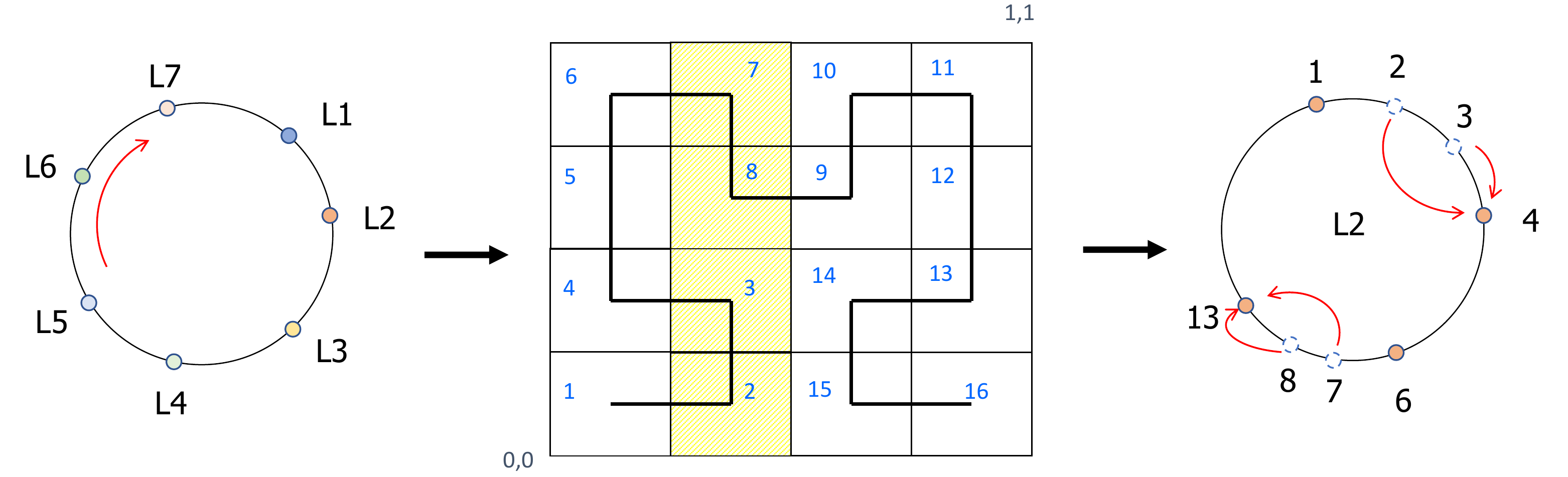}
\label{fig:multi}}
\end{center}
\caption{R-Pulsar space filling curve routing using simple and complex profiles.}
\label{fig:profiles}
\end{figure*}

\section{The R-Pulsar Framework}\label{sec:framework}
R-Pulsar is a lightweight memory mapped full-stack platform for supporting real-time data analytics across the cloud and edge in a uniform manner. 

The overall idea of R-Pulsar is to provide a platform for supporting real-time data analytics across the cloud and edge in a uniform manner by processing and managing data in the most optimal location. For this purpose, R-Pulsar relies on two concepts: Rendezvous Points (RP) and Profiles. 

An RP is the device performing streaming analytics, it can be a broadband access point, a forwarding node in a sensor network, or a server in a wired network.

R-Pulsar uses a profile-based abstraction called Associative Rendezvous programming abstraction (AR). The AR abstraction is used in R-Pulsar for the Pub/Sub subscription model and it is used to give users fine-grained control over the platform's runtime mechanisms and the deployment of streaming functions. A profile consists of keyword tuples made up of complete keywords, partial keywords, wildcards, and ranges. R-Pulsar consists of four layers: (1) a location-aware self-organizing overlay, (2) a content-based routing layer, (3) a memory-mapped  data-processing layer specially designed for high performance in small edge devices, and (4) a programming abstraction to allow users to express their stream processing applications in a dynamic data-driven fashion. The rest of these sections explain the internals of each layer. 
 
\subsection{Location aware overlay network layer}

The location-aware overlay network of R-Pulsar is motivated by two factors: first, the need to achieve low latencies and get rapid insights so decisions can be performed in a timely manner. The second one is due to traditional P2P overlay networks such as Chord~\cite{chord} or XOR~\cite{kademalia} suffering from high routing latencies and low efficiencies in data lookup due to the fact that they don't take location into consideration. For those reasons, we created a new location-aware, self-organizing, fault-tolerant peer-to-peer (P2P) overlay network, that enables an optimal placement of the analytics functions. 

R-Pulsar presents a new location-aware overlay network that exploits the use of quadtrees for avoiding the constant update of the routing tables. In addition, R-Pulsar overlay uses a 160bit unique identifier which allows it to connect more peers than you can address with IPv6. A point quadtree is a tree data structure in which each internal node has exactly four children. Each node represents a 2D bounded box covering a specific part of the space to index, using a root node to cover the entire area. 

A new RP is added to the system by determining which quadrant the RP point occupies, and inserting it to the quadtree from the root node to the appropriate leaf node. Every time the quadtree splits, the system creates four new P2P rings. Figure~\ref{fig:quadtree} is a graphical representation of the quadtree and the logical organization of the P2P network. During the join phase, messages are exchanged between a joining RP and the rest of the group. During this phase, the RP attempts to discover an already existing RP in the system to build its routing table. The joining RP sends a discovery message to the group. If the message is unanswered after a set duration (in the order of seconds), the RP assumes that it is the first in the system and it becomes the master RP of the ring. 

The master RP is responsible for manning the quadtree structure, and dictates when to divide P2P structure. Any time the overlay network is subdivided, the master RP randomly elects one of the RP nodes of the subdivision to be the master node of that region. If the master node of any of the regions fails, a new master RP election is performed using the Hirschberg and Sinclair algorithm, to know when the master peer is down peers send periodic keep alive messages. If the master peer doesn't respond to the keep alive the leader election is performed. Each RP master keeps a copy of the quadtree, so in the case of an RP failure the overlay network structure will never be lost. Also, every time an RP is added to the quadtree, the system ensures that each of the new four regions contain at least n amount of RP to guarantee a proper replication in an event of an RP failure. 

\subsection{Content based routing layer}\label{sec:frameworkc}

The content-based routing layer of R-Pulsar is motivated by the serverless model to keep track of where functions or data are stored in the overlay network. In addition, the content-based routing layer helps abstract the mapping between the user-specified profiles and map them in to a set of node overlay identifiers. It guarantees that all peers responsible for that profile will be found. 

The content-based routing layer uses the Hilbert Space Filling Curve (SFC)~\cite{SFC} to map the n-dimensional space to the one-dimensional space of the peer overlay. By applying the Hilbert mapping to this multi-dimensional space, each profile consisting of a simple keyword tuple can be mapped to a point on the SFC. Moreover, any complex keyword tuple can be mapped to regions in the keyword space and the corresponding clusters (segments of the curve). The one-dimensional index space generated corresponds to the one-dimensional identifier space used by the XOR overlay. Figure~\ref{fig:single} and~\ref{fig:multi} present a graphical representation of the routing process. Using SFCs, RP nodes corresponding to any simple or complex keyword tuple can be located. The routing process mainly requires three parameters: data, profile, and location. It differs depending on whether simple (Figure~\ref{fig:single}) or complex (Figure~\ref{fig:multi}) keywords are used.
\\\\
\textbf{Routing using simple keyword tuples:} The routing process consists of three steps. At first, the location of the RP node determines which of the $n$ overlay networks need to be reached. If the destination appears to be within another network, the message is forwarded to the master of the current overlay network before using the quadtree structure to route it in the proper overlay network. Second, the SFC mapping is used to construct the index of the destination RP node from the simple keyword tuple, and finally, the overlay network lookup mechanism is used to route to the appropriate RP node in the overlay. 
\\\\
\textbf{Routing using complex keyword tuples:} The complex keyword tuple identifies a region in the keyword space, which in turn, corresponds to clusters of points in the index space. Using those clusters, the overlay network lookup mechanism is used to route to all the of the responsible RP nodes in the overlay.

\subsection{Memory-mapped data processing layer}
R-Pulsar is motivated by the need to achieve high-performance stream analytics applications on single board computers. State of the art streaming analytics pipelines are known to be particularly compute and I/O intensive, resulting in challenges to perform real-time analytics due to limited read and write performances when offloading the data on external devices.

The processing layer is divided into three sub-layers:   

\begin{enumerate}

\item The data collection layer, implemented as a memory-mapped queue, gathers data from variable sources and its stored and as a middleware between the consumers.    
\item The stream processing layer, which processes the data and performs computations on the collected data.   
\item The data storage layer where data is stored using a DHT and is available to be queried.

\end{enumerate}
\hfill
\subsubsection{Data collection layer}
\hfill\\\\
The data collection layer is responsible for aggregating different types of data generated, from multiple geographically distributed sensors. Multiple messaging services are available such as Apache Kafka~\cite{kafka}, Google Pub/Sub~\cite{google}, Amazon Firehose~\cite{amazon}, or Mosquitto~\cite{mosquitto}. While some of those hubs are designed to be deployed on edge devices, they often offer limited performance for data collection and don't provide redundancy. To tackle these issues we designed and implemented a custom messaging hub specially designed for edge devices using a memory-mapped queue. Note that R-Pulsar memory-mapped queue offers the same guarantees as Mosquitto or Kafka (persistence, durability, and delivery guarantees). A memory-mapped file is a segment of virtual memory which has been assigned a direct correlation with some portion of a file. This file is physically present on-disk, which allows the operating system to ensure data access operations with better performance than standard file access. The core principle of a fast, while persistent, queue system like R-Pulsar is from an observation that sequential disk read is even faster than random memory read, as it can be seen in Table~\ref{tb:table}. The main advantages of the memory-mapped files are: (1) They are way faster than standard file access via normal I/O. (2) The operating system takes care of reading and writing to disk in the event of the program crashing.

\begin{table}[]
\scalebox{1.2}{
\begin{tabular}{|l|c|c|}
\hline
\multicolumn{1}{|c|}{\textbf{Operation}} & \textbf{Disk} & \textbf{RAM Memory} \\ \hline
Sequential read                          & 18.89 MB/s             & 631.34 MB/s \\ \hline
Sequential write                         & 7.12 MB/s              & 573.65 MB/s \\ \hline
Random read                              & 0.78 MB/s              & 65.96 MB/s              \\ \hline
Random write                             & 0.15 MB/s              & 65.88 MB/s              \\ \hline
\end{tabular}}
\centering

\caption{Measurments of Disk I/O vs RAM memory performance on a Raspberry Pi.} \label{tb:table}
\vspace{-3ex}
\end{table}

\hfill
\subsubsection{Stream processing engine}
\hfill\vspace{1ex}

This layer is in charge of transforming raw data stream into useful information and gather insights using a sequence of small processing units. R-Pulsar allows the end user to integrate any distributed online big data-processing system using customizable modules and generic functions. 

Also, some functionalities are pre-tailored for specific streaming engines: the current release of R-Pulsar has been validated using on-demand topologies (scaling up or down) designed for Apache Storm. Support for other stream processing engines is under development.
\\
\subsubsection{Memory mapped data storage layer}
\hfill\vspace{1ex}

Streaming analytics architectures deployed in the core of the network offer data replication across several nodes. We achieved a similar mechanism at the edge of the network by implementing a DHT that uses the overlay P2P network to automatically replicate the data and store using multiple RP located in same region. It guarantees that in the event of a RP crashing, the data will remain in the system and be seamlessly ready for queries. The storage layer relies on RocksDB~\cite{rocks}, an embedded database optimized for fast and low latency storage. RocksDB is optimized for datasets that are bigger than main memory. The database will keep the most recently used data in main memory, and it will store the least recently used data to disk. The database will write all index values to disk, too, when writing to disk RocksDB exploits the full potential of high read/write rates offered by flash and high-speed disk drives.

\subsection{Programming abstraction layer}\label{sec:programming}
The programming abstraction layer enables the interactions with the end user, and offers the ability to support the serverless streaming model. In this section we present two programming abstractions: the first abstraction is the AR programming abstraction, the second one is a rule-based system to specify functional rules that are triggered by data content, and determine which topologies are executed and where (e.g., at core and/or edge resources).
\\
\subsubsection{Associative Rendezvous abstraction}
\hfill\vspace{1ex}

\begin{figure}[htb!]
\begin{center}
\subfloat[]{\includegraphics[width=0.45\columnwidth]{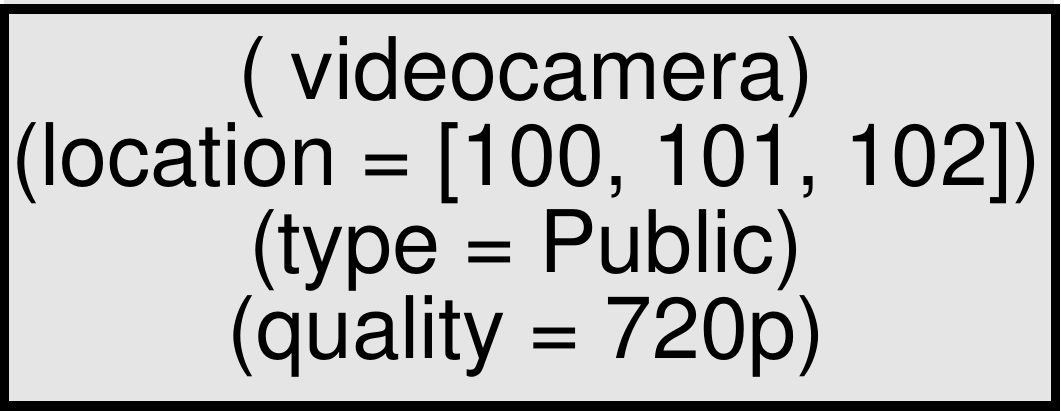}
\label{fig:data_profile}}
\subfloat[]{\includegraphics[width=0.45\columnwidth]{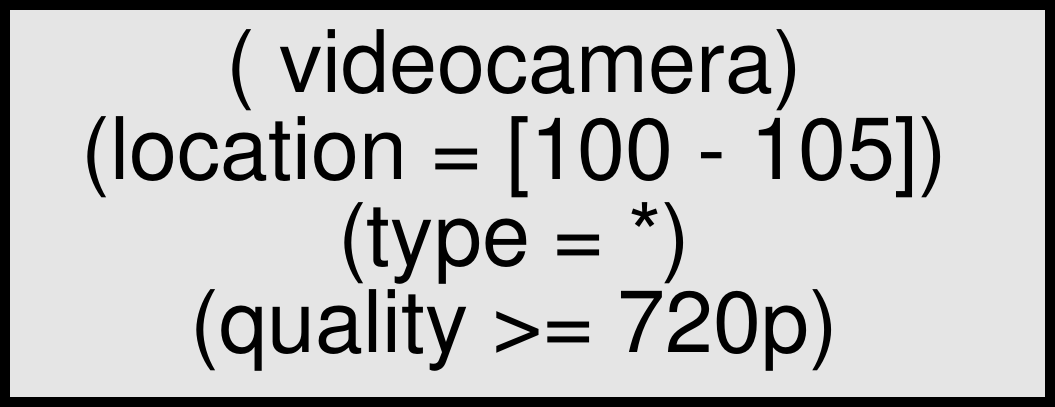}
\label{fig:interest_profile}}
\end{center}
\caption{Sample resource profiles: (a) a data resource profile for a sensor; (b) an interest resource profile for a client.}
\label{fig:samples_profiles}
\end{figure}

AR is a paradigm for content-based decoupled interactions with programmable reactive behaviors. Rendezvous-based interactions provide a mechanism for decoupling senders and receivers. These rendezvous interactions occurs at the RP nodes. Senders send messages to an RP without knowing from whom or where the receivers are. Similarly, receivers receive messages from an RP without knowing from whom or where the senders are. Note that senders and receivers may be decoupled in both space and time. The AR interaction model consists of three elements: messages, associative selection, and reactive behaviors, which are described below.
\\\\
\textbf{The AR message} is defined as the quintuplet: (header, action, data, location, topology). The data, location, and topology fields may be empty or contain a message payload with the location of the user. The action field of a message defines its reactive behavior when a matching occurs at a rendezvous point. The header includes the semantic profile with the credentials of the sender. A profile is a set of attributes and/or attribute-value pairs that define not only properties but also the recipients of the message. Attribute fields must be keywords from a defined information space, while value fields may be keywords, partial keywords, wildcards, or ranges from the same space. At rendezvous points, profiles are classified as resource profiles or as function profiles, depending on the action field of its associated message. A sample resource data profile used by a sensor to publish data is shown in Figure~\ref{fig:data_profile}, and a matching interest resource profile is shown in Figure~\ref{fig:interest_profile}.

\noindent
\textbf{The action field} of a message defines its reactive behavior at a rendezvous point. Basic reactive behaviors currently supported include {\it store, statistics, store\_function, start\_function, notify\_interest, notify\_data}, and {\it delete}.

The {\it store} action stores data in the appropriated rendezvous point DHT. The {\it statistics} action queries the system to retrieve detailed information about the characteristics and the status of the computational resources. The {\it store\_function} action allows users to submit and store used defined data analytics functions into the RPs, allowing us to have a distributed store where users can share and discover existing functions previously uploaded by other users. This avoids the need for rewriting the same function multiple times and facilitates the reproducibility of experiments. The {\it start\_function} allows users to trigger stream processing analytics on demand. It also causes the function profile to be matched against existing function profiles and if there is a match the function is executed. The {\it stop\_function} allows users to stop already running functions. Note that start\_function, store\_function and stop\_function are used for defining actions on function profiles, the rest of the actions are for defining the resource profiles.

The {\it notify\_interest} is used by data producers (IoT sensors) who want to be notified when there is interest in the data they produce, so that they can start sending data. The {\it notify\_data} is used by data consumers (functions or user-defined data analytics) who want to be notified when data matching their interest profile is stored in the system. The {\it delete} action deletes all matching profiles from the system.
\\\\
\textbf{Associative selection} is the content-based resolution and matching of profiles based on keywords (e.g. exact keywords, wildcards, partial wildcard, and ranges). A profile $p$ is a propositional expression. The elements in the profile can be an attribute, $a_i$ , or an attribute-value pair ($a_i$ , $v_i$), where $a_i$ is a keyword and $v_i$ may be a keyword, partial keyword, wildcard, or range. The singleton attribute $a_i$ evaluates to true with respect to a profile $p$ if and only if $p$ contains the attribute $a_i$ . The attribute-value pair ($a_i$ , $u_i$) evaluates to true with respect to a profile $p$ if and only if $p$ contains an attribute $a_i$ and the corresponding value $v_i$ satisfies $u_i$.

The AR interaction model defines three primitives: post, push, and pull. \textit{post(msg)}: The post primitive resolves the profile of the message and delivers the message to all the relevant rendezvous points. The profile resolution guarantees that all rendezvous points that match the profile will be identified. Nonetheless, the actual delivery relies on existing transport protocols. The post primitive uses the content-based routing and the location-aware overlay network layers to route the message to all the RPs. The other primitives are used for the data streaming part: \textit{push(peer, msg)} starts the streaming of data to a specific RP and \textit{pull(peer, msg)} enables the data to be consumed.
\\
\subsubsection{Data-driven decisions abstraction}\label{sec:programming-data}
\hfill\\\\
The data-driven abstraction uses a rule-based system, which contains all of the appropriate knowledge encoded into a set of IF-THEN rules. The system examines all the rule conditions (IF) and determines a subset, the conflict set, of the rules whose conditions are satisfied based on the data tuples. Out of this conflict set, one of those rules is triggered (fired). When a rule is fired, the action specified in its THEN clause is carried out. The loop for firing rules executes until one of two conditions is met: there are no more rules whose conditions are satisfied or a rule is fired. 

Progammers are allowed to specify two different types of rules. One lets you express data quality requirements which impose time constraints on the processing of the tuples, allowing the specification of a trade-off between the data quality and computational complexity. The second one allows to express content-driven rules which complement the data quality requirements by triggering further stream-processing topologies (on demand) either at the core or at the edge of the network if the data needs further processing due to quality of the data. We illustrate the behaviors and design in our previous paper~\cite{rules}.
\\
\subsubsection{API Examples}\label{sec:programming-api}
\hfill\vspace{1ex}

We will now consider two sets of API examples based on the disaster recovery use case presented in Section~\ref{sec:background}. The first set of examples illustrates the basics of the resource profiles for sensor registration, discovery, and acquisition. It depicts the code for required for processing LiDAR images from a drone. The second set of examples use the function profiles in order to store and trigger streaming topologies in R-Pulsar.

The first example is a resource profile for any sensor ready to join the R-Pulsar network. This profile needs to be declared prior to any data streaming. In Listing~\ref{lst:sensor}, line 1 shows the resource profile of a drone with a camera that can stream LiDAR pictures. On line 2, the ARMessage is created and the action is set to notify if interested which means that the sensor will not stream its data until someone declares interest in its data. On Line 3, the message is forwarded using the profile, note that in all of the examples to forward the profile the {\it post()} primitive is used  to translate the profile into a single or a collection of RPs. This way, the end-user never has to specify an IP address or a server, resulting in the entire resource pool being abstracted by a user-specified profile. 
\\
\begin{lstlisting}[language=mylang, caption={Data producer resource profile sample code.}, captionpos=b, label={lst:sensor}]
1 ARMessage.Header.Profile profile = 
  ARMessage.Profile.newBuilder()
  .addSingle("Drone").addSingle("LiDAR");
2 ARMessage msg = ARMessage.newBuilder()
  .setAction(ARMessage.NOTIFY_INTEREST)
  .setLatitude(40.0583).setLongitude(-74.4056))
  .setHeader(h).build();
3 producer.post(msg);
\end{lstlisting}

The profile on Listing~\ref{lst:ed} declares an interest in consuming data. 
R-Pulsar facilitates the sensor discovery and data acquisition by describing a profile that will be delivered to all matching sensors. 
Listing~\ref{lst:ed} illustrates the consumer profile, interested in consuming any LiDAR sensor data that matches the profile "Drone" and "Li*" (line 1), located in the specified range (40*, 70*). 
The device running the producer profile (from Listing 1) will be notified that there is a consumer interested in consuming its data, so the sensor will start streaming.
On Line 2, the ARMessage is created and attached to the profile created on line 1. 
The message is then forwarded using the profile so the drone can be notified and data can be streamed (Line 3).
\\
\begin{lstlisting}[language=mylang, caption={Data consumer resource profile sample code.}, captionpos=b, label={lst:ed}]
1 ARMessage.Header.Profile profile =
  ARMessage.Profile.newBuilder()
  .addSingle("Drone").addSingle("Li*")
  .addSingle("lat:40*")
  .addSingle("long:-74*").build();
2 ARMessage msg = ARMessage.newBuilder()
  .setHeader(h).setAction
  (ARMessage.NOTIFY_DATA).build();
3 producer.post(msg);
\end{lstlisting}

The profiles can be used in two ways: for discovering/subscribing to data publishers for deploying functions across the edge and the cloud. Listing~\ref{lst:ed2} illustrates a function profile to store the post\_processing\_func in the system (line 1).  
On Line 2, the ARMessage is created and attached to the profile created on line 1.  
The message is then forwarded using the profile, and the function is stored in all the responsible RPs (Line 3).
\hfill\\
\begin{lstlisting}[language=mylang, caption={Store processing function as post\_processing\_function in the R-Pulsar overlay network.}, captionpos=b, label={lst:ed2}]
1 ARMessage.Header.Profile profile = 
  ARMessage.Profile.newBuilder()
  .addSingle("post_processing_func").build();
2 ARMessage msg = ARMessage.newBuilder()
  .setHeader(h)
  .setAction(ARMessage.STORE_FUNCTION).build();
3 producer.post(msg);
\end{lstlisting}

Consequently, a profile and a decision (IF-THEN rule) can be created to trigger the previous function (post\_processing\_func) when a condition is met. 
On Listing~\ref{lst:t-rule}, the resulting action is created and attached to the profile from Listing~\ref{lst:t-start2} (Line 1), which will be sent when the rule is satisfied.

In addition, on Listing~\ref{lst:t-rule}, Line 2 defines a rule that will be constantly evaluated for every data element.
If the condition of this rule is met, the profile from Listing~\ref{lst:t-start2} (Line 1) will be forwarded, resulting in the execution (trigger) of the topology that previously stored. 

By combining the two abstractions implemented in R-Pulsar, the AR programming abstraction and the IF-THEN rule abstraction, developers can create on-demand data-driven pipelines over the edge and the cloud.

\begin{lstlisting}[language=mylang, caption={Data driven rule abstraction to trigger post\_processing\_function.}, captionpos=b, label={lst:t-rule}]
1 ActionDispatcher topo1 = new 
  TriggerTopologyReaction(T-profile);
2 Rule rule1 = new Rule.Builder()
  .withCondition("IF(RESULT >= 10)")
  .withConsequence(topo1)
  .withPriority(0).build();
\end{lstlisting}

\hfill

\begin{lstlisting}[language=mylang, caption={Profile to execute data streaming post\_processing\_function.}, captionpos=b, label={lst:t-start2}]
1 ARMessage.Header.Profile T-profile = 
  ARMessage.Profile.newBuilder()
  .addSingle("post_processing_func").build();
2 ARMessage msg = ARMessage.newBuilder()
  .setAction(ARMessage.START_FUNCTION})
  .setHeader(h).build();
3 producer.post(msg);
\end{lstlisting}

\subsection{Implementation Overview}

The current implementation of R-Pulsar builds on a custom build of TomP2P~\cite{tom}. TomP2P is a distributed hash table which provides a decentralized key-value infrastructure for distributed applications. The underlying communication framework uses Java NIO~\cite{java} to handle many concurrent connections. The overall operation of the location-aware overlay network consists of two phases: bootstrap and running. 

During the bootstrap phase (or join phase), messages are exchanged between a joining RP and the rest of the group. During this phase, the joining RP attempts to discover RPs already existing in the system to build its routing table. The joining RP sends a discovery message to the group. If the message is unanswered after a set duration (in the order of seconds), the RP assumes that it is the first in the system. If an RP responds to the message, the joining RP and the rest of the RPs update their routing tables.

The running phase consists of stabilization and user modes. In the stabilization mode, an RP responds to queries issued by other RPs in the system. The purpose of the stabilization mode is to ensure that routing tables are up to date and to verify that other RPs in the system have not failed or left the system. In the user mode, each RP interacts at the programming abstraction layer. The programming abstraction layer matching engine at each RP is based on RocksDB, a key value database. Once the system is operating in user mode, RPs allow external entities to use the programming abstraction layer to communicate with each other to offer and request data and computation. These entities can include: a) users that might want to retrieve specific data or perform certain computation over data found using a query; b) IoT devices that can produce and consume data based on specific interests; and c) computational resources, such as data analytics and streaming platforms, clouds, or high-performance computing clusters that offer their computational capabilities. 

\section{Evaluation}\label{sec:evaluation}
This section presents the experimental evaluation of R-Pulsar. First, we present the performance of individual components of the stream processing pipeline (collection, processing, and serving layers) compared to established solutions. Second, we implemented an end-to-end data pipeline based on image processing and evaluated the response time of the application as a whole.

Experiments are evaluated using three different environments:   
\begin{itemize}
\item Raspberry Pi System: a Raspberry Pi 3 with 4x ARM Cortex-A53 1.2GHz, 1GB LPDDR2 of RAM and 10/100 Ethernet.
\item Android System: a Motorola Moto G5 Plus with a Qualcomm® Snapdragon™ 625 processor with 2.0 GHz octa-core CPU, 3GB of RAM
\item Cloud System: Chameleon cluster with 5 instances of type m1.small and 5 instances of type m1.medium to simulate computation capabilities of a Raspberry Pi and the hardware heterogeneity that IoT presents.
\end{itemize}

\subsection{Framework/System Performance}
\subsubsection{Performance of the messaging layer over Raspberry Pi system}
\hfill\\\\
The first experiment aims to evaluate the throughput and the throughput stability of three different messaging systems with four different message sizes. The ability to maintain a steady throughput is critical to systems that are communication-intensive such as stream processing engines. 

\begin{figure}[h]
  \includegraphics[width=0.5\textwidth]{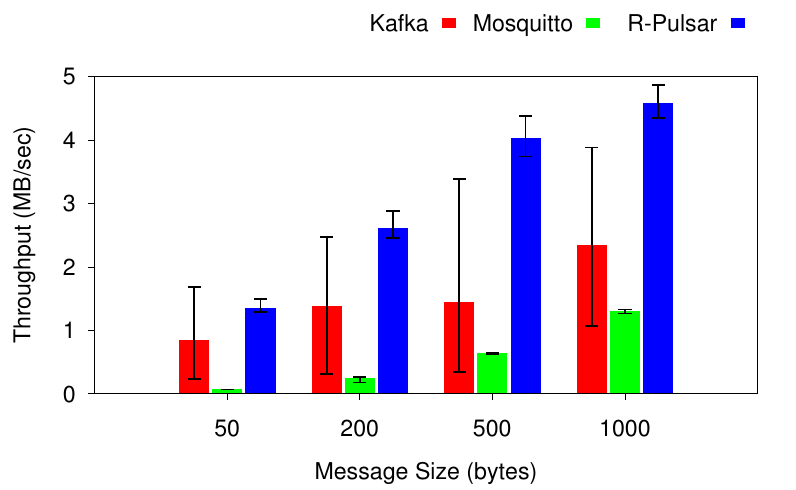}
  \caption{Single producer R-Pulsar throughput vs Kafka \& Mosquitto throughput.}
  \label{fig:ProducerBar}
\end{figure}

Figure~\ref{fig:ProducerBar} compares Apache Kafka, \textit{de facto} standard on the market, Mosquitto, a stack design for IoT and R-Pulsar, our proposed solution. These three distributed engines offer the same guarantees in terms of persistence, durability, and delivery – the main difference between them is the way to store information: Apache Kafka and Mosquitto store messages on disk while R-Pulsar stores them in the main memory. Results show that R-Pulsar pub/sub messaging system overperforms Kafka up to a factor of 3 and Mosquitto up to a factor of 7. Considering a traditional IoT scenario with small messages being streamed at a high rate of arrival, Kafka and Mosquitto perform poorly compared to R-Pulsar. We observed that Apache Kafka does not exhibit a constant throughput resulting in high variability of throughput performance. This is explained by the fact that Kafka continuously stores messages on disk overwhelming the file system and producing an unpredictable throughput. Since R-Pulsar uses a memory-mapped queue, not only does it produce a higher throughput, but it also is a more predictable and steady throughput, making it suitable to support real-time data analytics on single board devices.
\hfill\vspace{1ex}
\subsubsection{Performance of the query and store operations over Raspberry Pi system}
\hfill\vspace{1ex}

The next setup illustrates the performance of R-Pulsar's internal DHT compared to self-contained, embedded, lightweight data storage systems that offer the same guarantees.
We compare R-Pulsar next to lightweight SQL (SQLite) and non-SQL (NitriteDB) storage systems, the main types of data storage systems implemented in stream processing engines. In this context, the performance of such systems is defined by their ability to store and query/retrieve data.

\begin{figure}[h!]
  \includegraphics[width=0.5\textwidth]{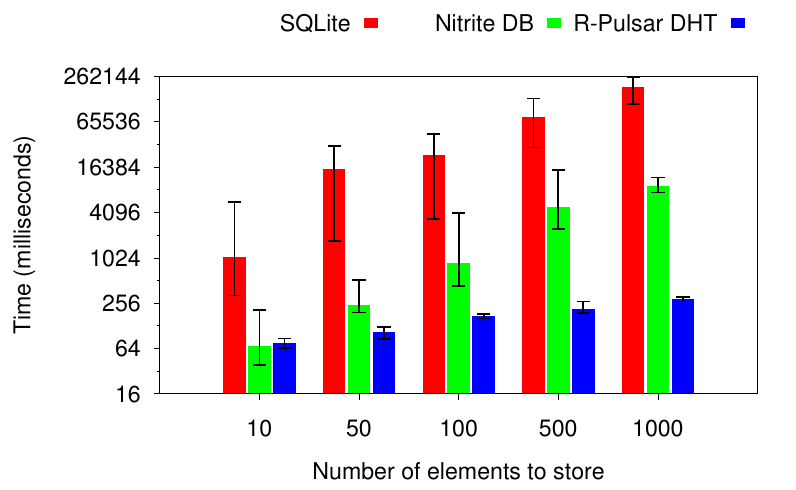}
  \caption{R-Pulsar DHT storage vs SQLite and Nitrite DB.}
  \label{fig:DBInsertBar}
\end{figure}

We present three different results in this setup. Figure~\ref{fig:DBInsertBar} shows that R-Pulsar outperforms the best solution (SQLite) by a factor of 32 when storing elements. This is due to the fact that SQLite and NitriteDB store all records on disk, and they are not optimized for fast, low-latency storage such as flash drives or high-speed disk drives. R-Pulsar uses a combination of in-memory and disk to get higher throughput in single board devices. The storage system keeps the most recently used data in main memory and stores the least recently used data to disk.

\begin{figure}[h!]
  \includegraphics[width=0.5\textwidth]{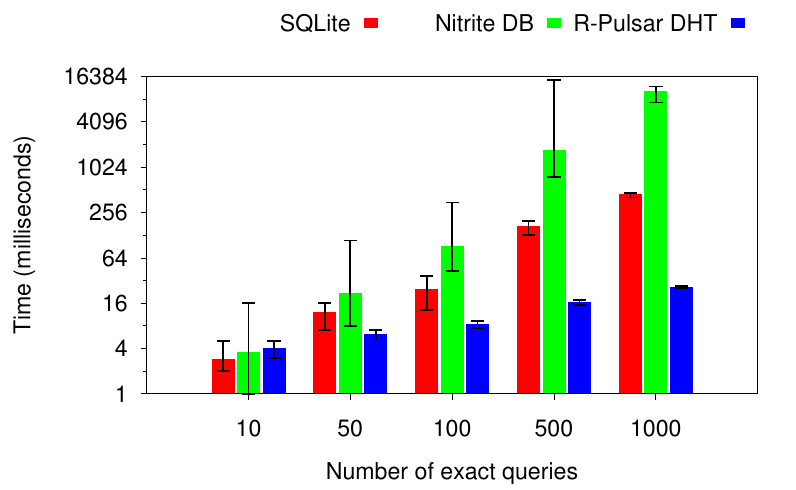}
  \caption{R-Pulsar exact query layer vs SQLite and Nitrite DB.}
  \label{fig:DBExactBar}
\end{figure}

Figure~\ref{fig:DBExactBar} presents the case for exact queries. Exact queries are defined by exact keywords and return a single result rather than multiple ones.

\begin{figure}[h!]
  \includegraphics[width=0.5\textwidth]{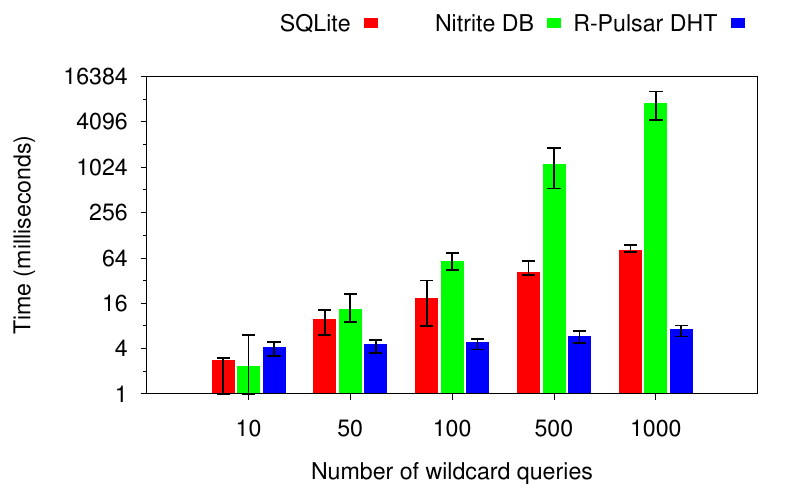}
  \caption{R-Pulsar wildcard query layer vs MySQLite and Nitrite DB.}
  \label{fig:DBWildBar}
\end{figure}

Similarly, Figure~\ref{fig:DBWildBar} presents the comparison of the same system using wildcard queries, which are defined by wildcards that match any character sequence and may return multiple results rather than a single one. 

Those three experiments show that, while Nitrite DB and SQLite are both slightly faster for small workloads, R-Pulsar shows better performance as the workload increases.

\hfill\vspace{1ex}
\subsubsection{Performance of the messaging layer over Android system}
\hfill\vspace{1ex}

As of 2018, Android is the leading mobile operating system ~\cite{market}, and is integrated to a large amount of devices, including IoT systems. In this context, an implementation of R-Pulsar on Android phones has been developed to showcase its ability to be deployed on such devices.

\begin{figure}[h!]
  \includegraphics[width=0.5\textwidth]{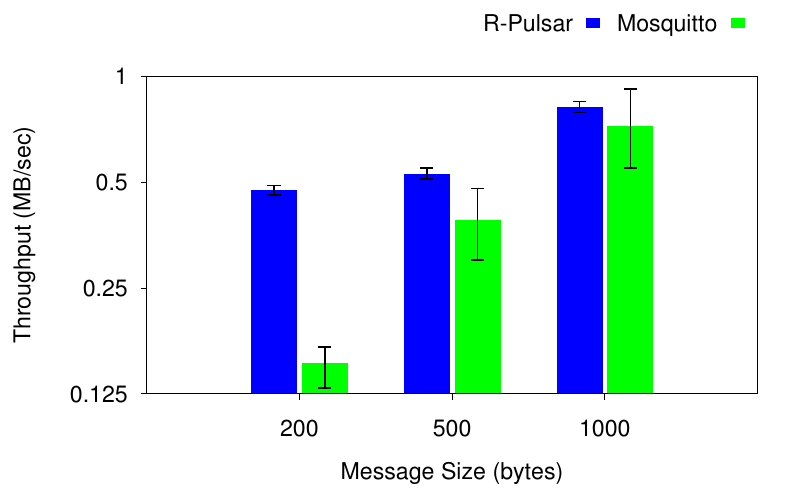}
  \caption{Single Producer R-Pulsar throughput vs Mosquitto throughput on Android Phone}
  \label{fig:ProducerPhone}
\end{figure}

Figure~\ref{fig:ProducerPhone} compares the throughput of R-Pulsar and Mosquitto, an open-source client of MQTT and MQTT-SN messaging protocols aimed at existing and emerging applications for the Internet of Things (IoT)~\cite{paho}. Experiment setup describes a single Android device as a data producer and a Rasberry Pi as Rendezvous Point (RP). Those two solutions are designed to be deployed in constrained devices with the exact same delivery guarantees. On average, R-Pulsar performs better than Mosquitto by a factor of $\sim$10, specifically for small messages. Moreover, Mosquitto presents a larger variability in terms of performance as some of the results are above R-Pulsar's numbers. The variability is due to the fact that Mosquitto also uses disk to store messages and ends up overwhelming the file system.

\hfill\vspace{1ex}
\subsubsection{Routing overhead over Android and Raspberry pi systems}
\hfill\vspace{1ex}

This set of examples is designed to show the routing overhead associated with R-Pulsar when deployed on Android and Raspberry Pi systems. The evaluation is performed by simulating the need to store or retrieve portions of data as the number of RPs on a given region grows and as the complexity of the profile to route increases. The profile complexity is defined by its number of properties, for example, a 2D profile is composed of two properties such as type and location. 

\begin{figure}[h]
  \includegraphics[width=0.5\textwidth]{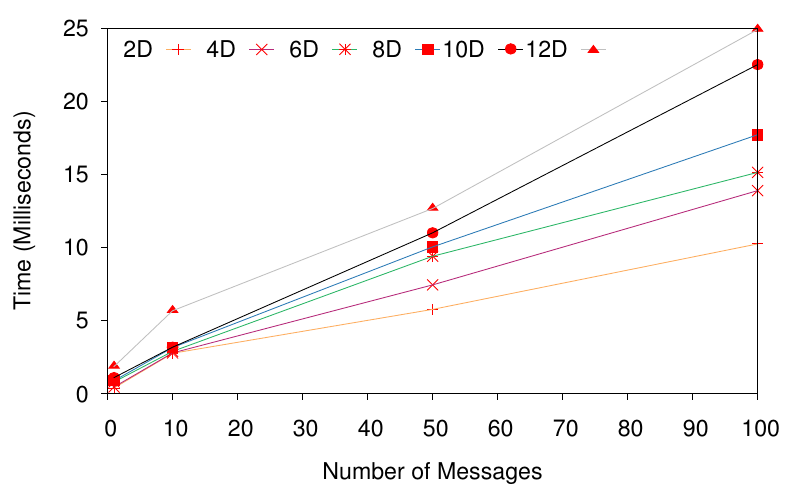}
  \caption{R-Pulsar space filling curve routing overhead and scalability on Android Phone}
  \label{fig:Phone}
\end{figure}

\begin{figure}[h]
  \includegraphics[width=0.5\textwidth]{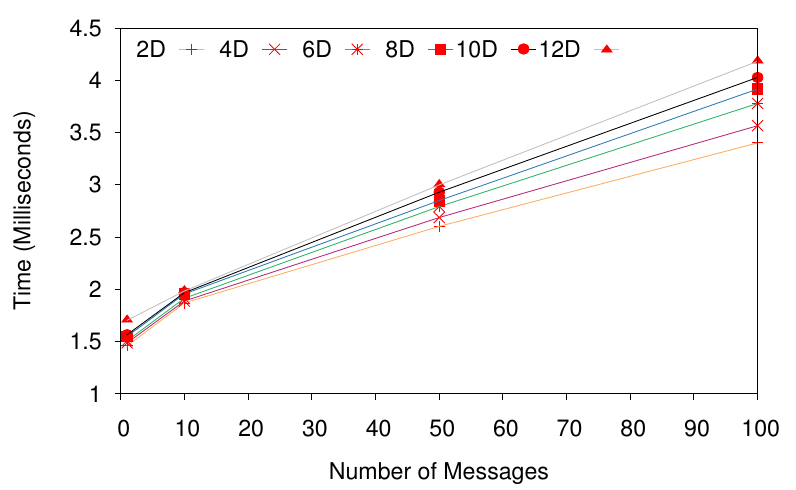}
  \caption{R-Pulsar space filling curve routing overhead and scalability on Raspberry Pi}
  \label{fig:Raspberri}
\end{figure}

Figure~\ref{fig:Phone} shows that, when the profile complexity increases by a factor of 6, the time required to route messages increases by 2.5. Similarly, when the system increases the number of messages to send by a factor of 100, the time required to route one message increases by $\sim$25, showing that the routing overhead scales efficiently in both cases, when the messages are becoming more and more complex and as the number of messages required to send increases on an Android phone. On the other hand, Figure~\ref{fig:Raspberri} shows that when the profile complexity increases by a factor of 6, the time required to route messages increases by $\sim$1.2. Likewise, when the system increases the number of messages that needs to be sent by a factor of 100, the time required to route one message increases by $\sim$2.5, showing that the routing overhead scales even more efficiently on a Raspberry Pi than on an Android phone.

\hfill\vspace{1ex}
\subsubsection{Scalability of store/query operation on Chameleon cluster}
\hfill\vspace{1ex}

In addition to deploying R-Pulsar on Raspberry Pis and Android systems, we deployed it on virtual machines using the Chameleon Cluster, a configurable experimental environment for large-scale cloud research~\cite{chameleon}. These experiments aim at stressing the system and evaluate the storage and query scalability of R-Pulsar using multiple workload sizes. The workloads used for these tests are the following: Workload 1 (W1) stores/queries one element, Workload 2 (W2) stores/queries 10 different elements, Workload 3 (W3) stores/queries 50 different elements, and Workload 4 (W4) stores/queries 100 different elements. For this test, all RP nodes are part of the same P2P network and same geographical region in order to evaluate how R-Pulsar will scale to the number of RP increases in each region. 

\begin{figure}[h!]
  \includegraphics[width=0.5\textwidth]{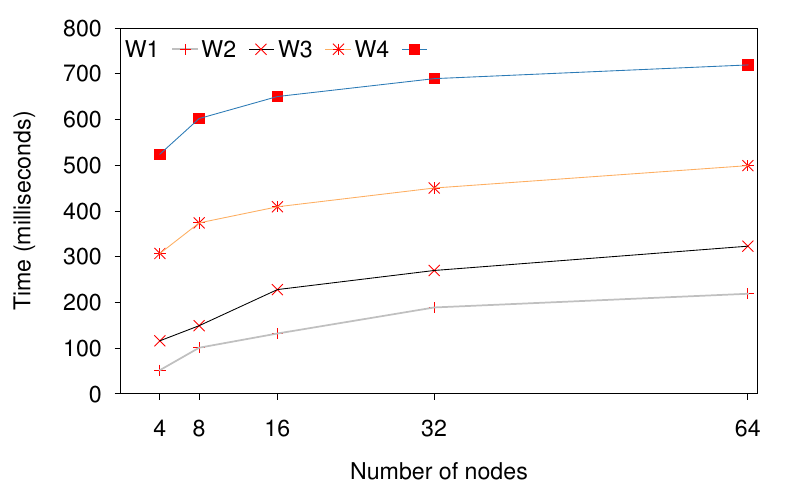}
  \caption{R-Pulsar store query scalability test on Chameleon cloud.}
  \label{fig:ProducerLine}
\end{figure}

Figure~\ref{fig:ProducerLine} presents the scalability evaluation of R-Pulsar store operation. It shows that for storing a single element (w1), the runtime increases by a factor of \~4 and the system size increases by a factor of 16 (from 4 nodes to 64 nodes). As the system expands, the number of intermediary nodes involved in routing the query grows, causing an increase in the runtime. The storage of 100 different elements (w4) forces the system to store elements into multiple destinations, and once again, the rate of increase of message runtime is smaller than that of the system size. 

\begin{figure}[h!]
  \includegraphics[width=0.5\textwidth]{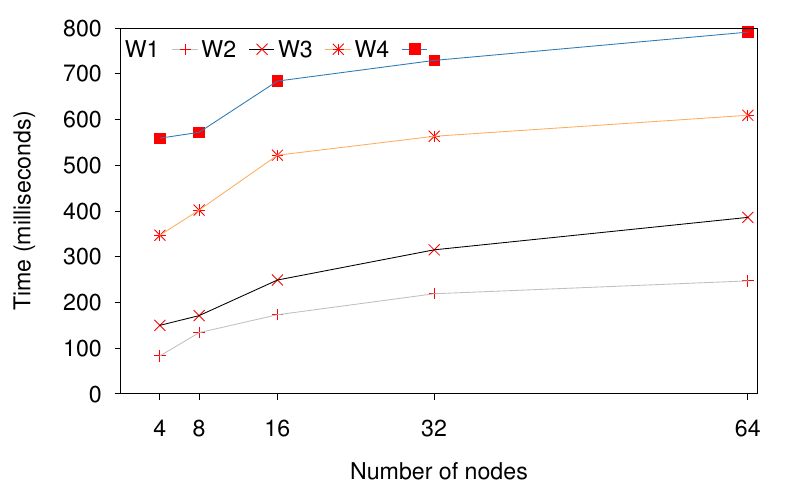}
  \caption{Exact query scalability test on Chameleon cloud.}
  \label{fig:ProducerLineEx}
\end{figure}

Figure~\ref{fig:ProducerLineEx} presents an experiment similar to the previous one, however in this case, the exact query scalability is tested. It shows that for the query of a single element (w1), the runtime increases by a factor of 2.8 when the system size increases by a factor of 16 (from 4 nodes to 64 nodes).

\subsection{Use case Evaluation}

\begin{figure}[h!]
  \centering
    \includegraphics[width=0.9\columnwidth]{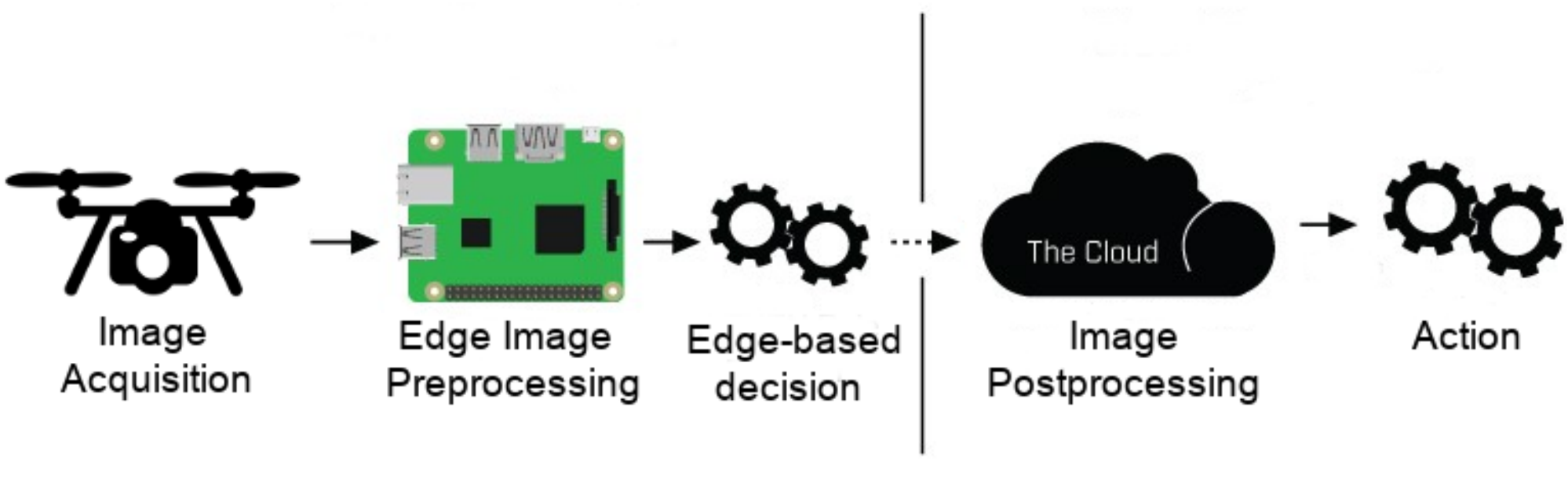}
  \caption{Disaster recovery workflow physical architecture.} \label{fig:workflow}
\end{figure}

In this section, experiments apply the motivating use case of disaster response workflow described in Section \ref{sec:background}. 
Figure~\ref{fig:workflow} depicts the setup used for this test. A mobile device (drone) flying over affected areas captures LiDAR images. It locally processes images on a Raspberry Pi and determines if they need to be sent to the core of the network for further processing or stored on the edge for fast access. 

The drone is represented as a data producer by the means of a Raspberry Pi that continuously emits LiDAR images and processes them through the data analytics pipeline. 

The ability to pre-process data and trigger reactive behavior is implemented using the rule-based system described in Section~\ref{sec:programming-data}. In this context, the drone flies around affected areas and captures LiDAR images and pre-processes the data \textit{in situ}. At the end of the computation, every images are evaluated and if an image needs further processing it is sent to the cloud. To implement it, portion of the code displayed in Section~\ref{sec:programming-api} was used. 

Following this setup, we compared R-Pulsar with two data analytics pipelines (data collection + analytics + storage):  
\begin{itemize}
\item Apache Kafka + Apache edgent + SQLite
\item Apache Kafka + Apache edgent +  Nitrite DB
\end{itemize}

\begin{figure}[h!]
  \includegraphics[width=0.5\textwidth]{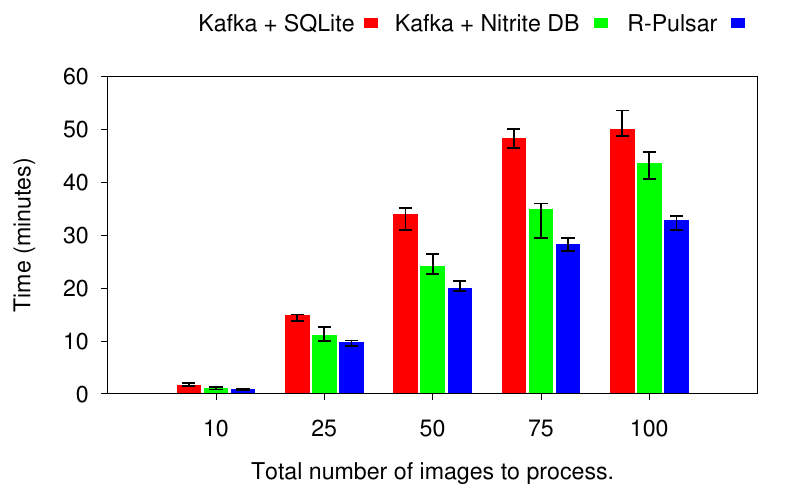}
  \caption{End to end performance test using a disaster recovery workflow on a Raspberry Pi.}
  \label{fig:EndToEnd}
\end{figure}

Figure \ref{fig:EndToEnd} shows that by using R-Pulsar software stack, we observe a gain in response time up to 36\% compared to traditional stream processing pipelines. Being able to obtain our results 36\% faster with R-Pulsar means that affected areas can go back to normal much faster and we can cover much larger areas in less amount of time. R-Pulsar is not only limited to this workflow, it can be applied to any edge analytics workflow that requires fast data insights. 

\section{Conclusion \& Future work}\label{sec:conclusion}

Despite high computing power of clouds and high network speeds between the network's edge and core, there is an emerging gradual recognition to move some parts of the computation to the edge of the network.
In this context, we propose R-Pulsar as an extension to the traditional stream processing model, which facilitates real-time data analytics between the cloud and the edge in a uniform manner. 
R-Pulsar relies on a profile-based paradigm to abstract heterogeneous edge and core resources to enable easier and more intuitive development of real-time analytics and data pipelines. 

In this technical report, we describe the requirements and use case, as well as the architecture and implementation of R-Pulsar using Raspberry Pi, Android, and cloud systems. 
Experiments described an extensive evaluation both at system- and application- level. 
Authors individually evaluated components of the software stack: throughput with increasing message size, query/store operations performance, and routing overhead as the complexity/dimension of profiles increase.
Beyond the scalability, authors also compared the end-to-end performance of data pipelines when dealing with stream processing of images on a real-life use case. Results show a 36\% speed up when tested using a real-life scenario compared to other established data analytics solutions.

We believe that R-pulsar is suitable for online stream processing both at the edge and the core of the network, thanks to its lightweight design and steady performance. The latest version of R-Pulsar is available on GitHub~\cite{pulsar} under Apache License 2.0.

As for future work, we are currently working towards extending R-Pulsar to support advanced storage strategies, with regard to cost of data movements and energy efficiency. Furthermore, the system could be adapted to replicate data based on query trends over federations of nodes. 

%\section{Acknowledgements}
%This work is supported in part by NSF via grants numbers ACI 1339036, ACI 1441376. The research was conducted at  the Rutgers Discovery Informatics Institute (RDI$^2$).

\bibliographystyle{unsrt}
\bibliography{Bibliography-HPDC}

\end{document}